\documentclass{PoS}

\usepackage{url}
\newcommand{\be}{\begin{equation}}
\newcommand{\ee}{\end{equation}}
\newcommand{\bea}{\begin{eqnarray}}
\newcommand{\eea}{\end{eqnarray}}
\newcommand{\bi}{\begin{itemize}}
\newcommand{\ei}{\end{itemize}}
\newcommand{\ben}{\begin{enumerate}}
\newcommand{\een}{\end{enumerate}}

\newcommand{\lp}{\left(}
\newcommand{\rp}{\right)}
\newcommand{\gsim}{\lower.7ex\hbox{$\;\stackrel{\textstyle>}{\sim}\;$}}
\newcommand{\lsim}{\lower.7ex\hbox{$\;\stackrel{\textstyle<}{\sim}\;$}}

\title{Parton Distributions in the Higgs Boson Era}

\ShortTitle{Parton Distributions in the Higgs Boson Era}

\author{\speaker{Juan Rojo}\thanks{This work is supported by a Marie Curie 
Intra--European Fellowship of the European Community's 7th Framework Programme under contract number PIEF-GA-2010-272515. }\\
        PH Department, TH Unit, CERN, CH-1211 Geneva 23, Switzerland\\
        E-mail: \email{juan.rojo@cern.ch}}

\abstract{Parton distributions are an essential ingredient of the LHC program.
PDFs are
relevant for precision Standard Model measurements, for Higgs boson
characterization as well as for New Physics searches.
In this contribution I 
review recent progress  in the determination of the parton
distributions of the proton during the last year.
Important developments 
include the impact of new LHC measurements
to pin down poorly known PDFs, studies of
theoretical uncertainties, 
higher order calculations for processes relevant for PDF
determinations,
 PDF benchmarking exercises with LHC data, as well as
methodological and statistical improvements in
the global analysis framework.
I conclude with some speculative considerations about future directions
in PDF determinations from the theory point of view.}

\FullConference{XXI International Workshop on Deep-Inelastic Scattering and Related Subjects -DIS2013,\\
		22-26 April 2013\\
		Marseille, France}

\begin{document}

\paragraph{Parton distributions and LHC phenomenology.}

\begin{figure}[h]
\centering
\includegraphics[scale=0.22]{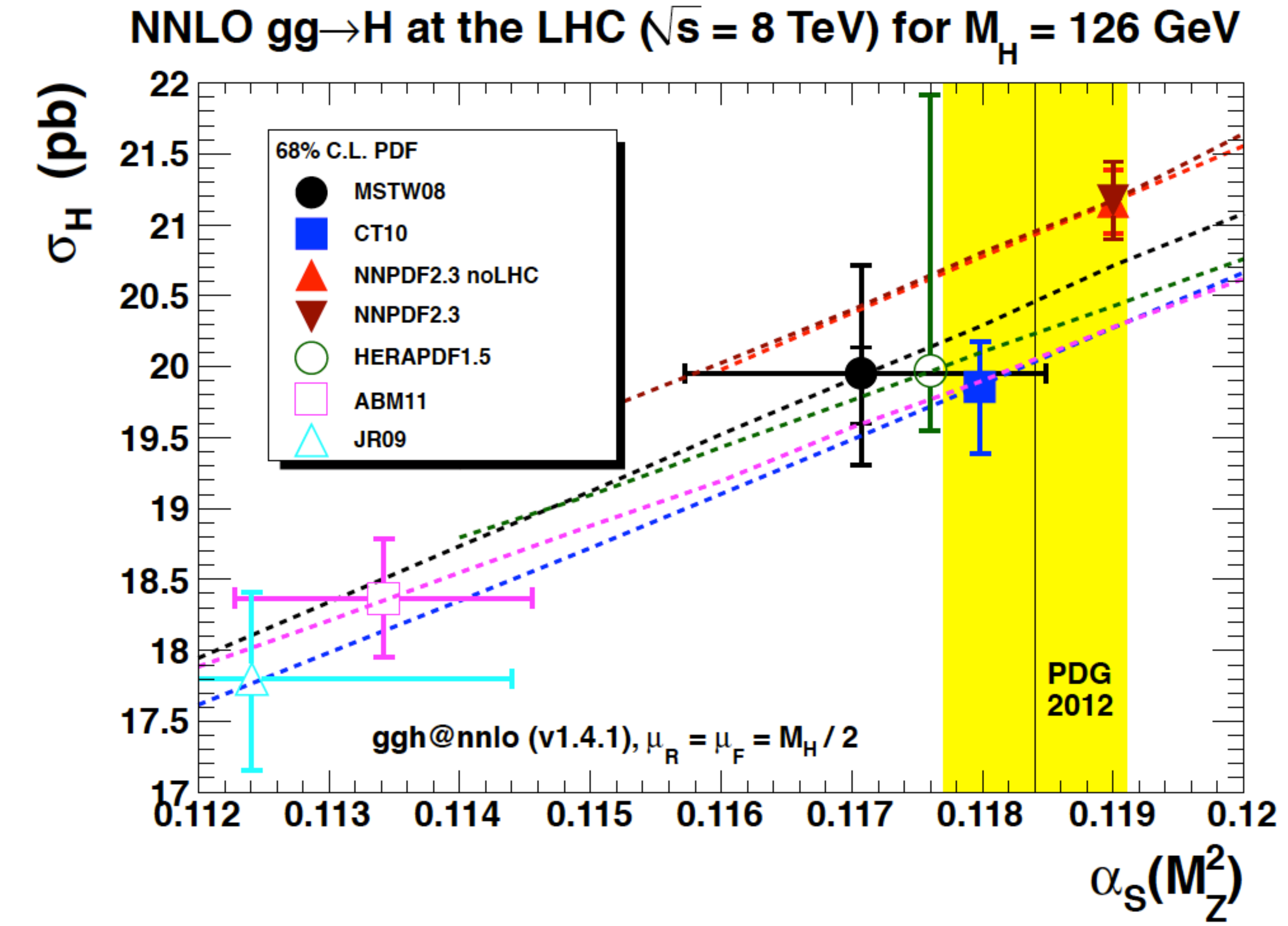}
\includegraphics[scale=0.14]{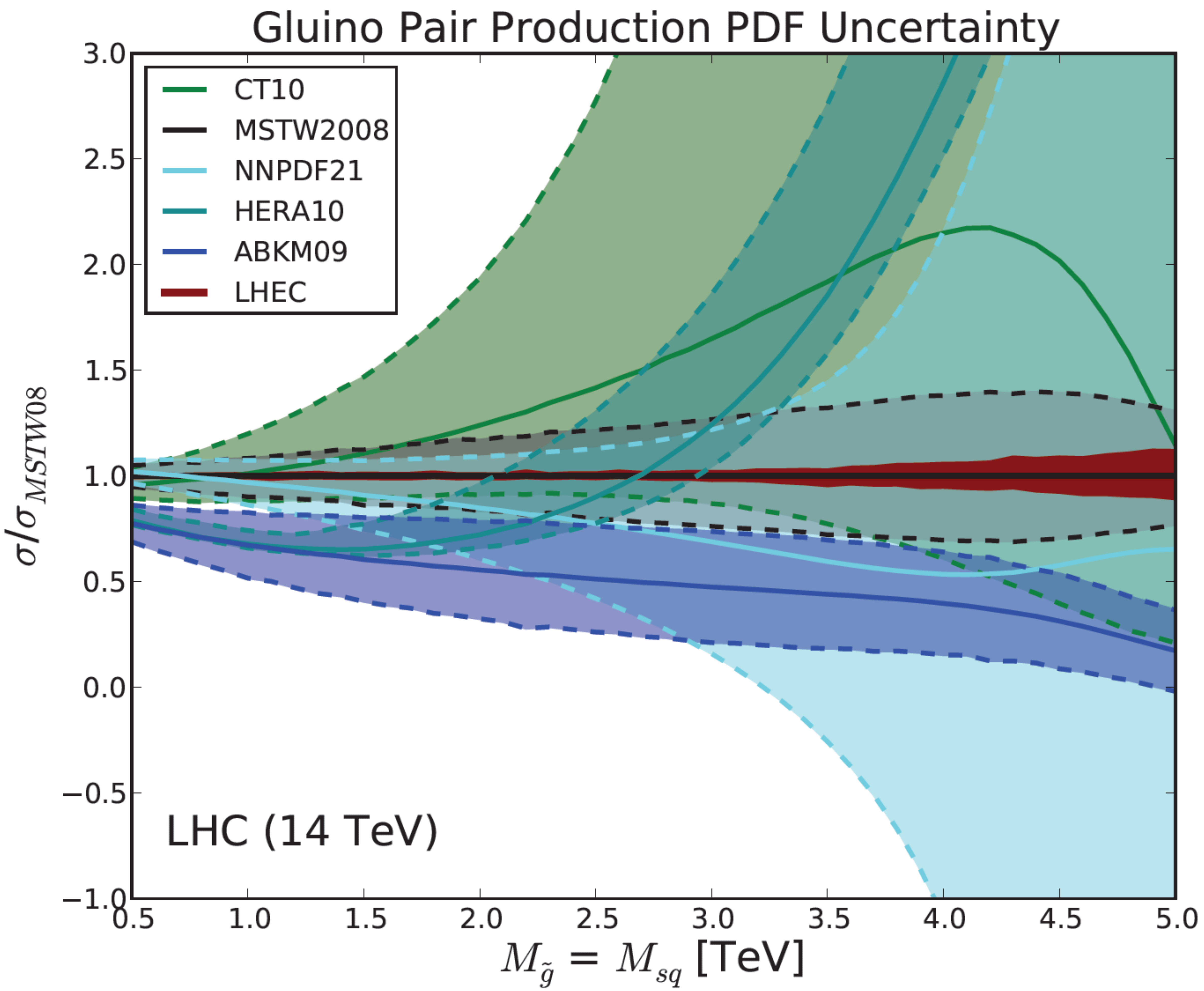}
\caption{\small  (left) PDF dependence of the Higgs production
cross section in gluon fusion at 8 TeV~\cite{Forte:2013wc}. (right)
PDF uncertainties for the production of heavy supersymmetric
particles at the LHC, from~\cite{AbelleiraFernandez:2012cc}.  
}
\label{fig:higgs}
\end{figure}

Precision physics at the LHC requires ever better  
estimates of PDF uncertainties~\cite{Forte:2013wc}. 
The relevance of PDFs for the LHC physics program
can be well illustrated by three representative examples.
 First of all, PDF uncertainties in
Higgs production are
 a fundamental limit  to the extraction the Higgs
boson couplings from experimental data, and thus degrade 
the prospects for Higgs
characterization~\cite{Dittmaier:2012vm,Watt:2011kp}.
The limiting factor is provided by the
spread of the predictions from recent NNLO PDF
sets for Higgs production in gluon fusion, shown in
Fig.~\ref{fig:higgs}. 
Second, PDFs at large momentum fractions $x$ 
suffer from substantial uncertainties, due to the lack
of experimental constraints.
These uncertainties translate into a lack of predictive power
for the production of very massive particles in the TeV range, expected
in most New Physics scenarios, such
as for heavy supersymmetric
 particles~\cite{AbelleiraFernandez:2012cc,Kramer:2012bx}, see
also Fig.~\ref{fig:higgs}.
The final example is provided by the determination of the
$W$ boson mass at the Tevatron and at the LHC: this precision
observable provides stringent consistency tests of the Standard
Model through the global electroweak fit~\cite{Baak:2012kk}, 
but is currently limited by 
PDF uncertainties at the Tevatron~\cite{Group:2012gb}, and will be even more
the case at the LHC~\cite{Bozzi:2011ww}. 
%

%%%%%%%%%%%%%%%%%%%%%%%%%%%%%%%%%%%%%%%%%%%%%%%%%%%%%%%%%%
%%%%%%%%%%%%%%%%%%%%%%%%%%%%%%%%%%%%%%%%%%%%%%%%%%%%%%%%%%

The aim of this contribution is to 
present a succinct overview of recent progress in PDF determinations
since DIS2012. More
 comprehensive recent reviews of PDF developments can be found
in Refs.~\cite{Forte:2013wc,DeRoeck:2011na,Perez:2012um}.

\paragraph{PDF updates and benchmarking with LHC data.}

Various collaborations provide regular updates of their
PDF sets. The latest releases from each group are ABM11~\cite{Alekhin:2012ig}, 
CT10~\cite{Gao:2013xoa}, HERAPDF1.5~\cite{Radescu:2010zz,CooperSarkar:2011aa}, 
MSTW08~\cite{Martin:2009iq} and NNPDF2.3~\cite{Ball:2012cx}. 
All of these sets provide their PDFs both at NLO and NNLO, and a wide
range of values of the strong coupling $\alpha_s(M_Z)$ is available.
A recent addition to the family of PDF anaysis is the CTEQ-JLAB collaboration
with their new CJ12 set~\cite{Owens:2012bv}, 
which emphasizes the use of large-$x$ DIS data
to reduce the experimental PDF uncertainties and an accurate
treatment of higher twists and deuteron corrections.

\begin{figure}[h]
\centering
\includegraphics[scale=0.124]{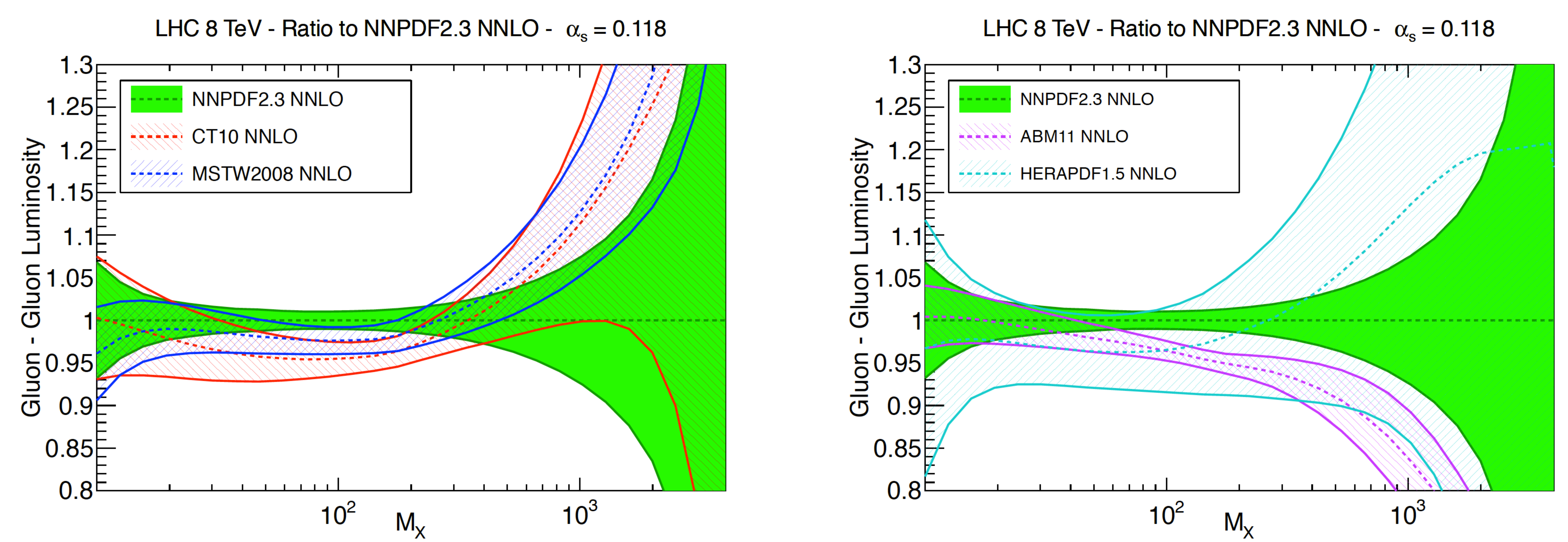}
\caption{\small Comparison of the gluon-gluon luminosities as
 a function of the final state mass $M_X$ for the five NNLO PDF
sets of the benchmark comparison from Ref.~\cite{Ball:2012wy}.
}
\label{fig:bench}
\end{figure}

A recent benchmarking exercise comparing NNLO PDFs between
them and with Tevatron and LHC data was presented in~\cite{Ball:2012wy}, 
updating previous benchmarking studies~\cite{Watt:2011kp}. 
The comparisons were performed at the level of PDFs,
parton luminosities, LHC 8 TeV total cross sections and differential
distributions for jet and $W,Z$ data from the Tevatron and LHC 7 TeV.
A detailed discussion was presented concerning
the quantification of data versus theory agreement though the use
of various $\chi^2$ estimators. %
The main outcome of the comparisons was the confirmation
of the reasonable agreement between CT, MSTW and NNPDF, 
already observed in previous benchmarks. 
The agreement improves if 2010 NLO PDFs are compared to 2012 NNLO 
PDFs for a variety of important observables, with the unfortunate exception
of gluon-gluon Higgs production for $m_H=$ 125 GeV.
The
HERAPDF set leads to similar central values as the global sets but
larger PDF uncertainties due to the reduced dataset, while 
ABM leads to softer large-$x$ gluons and harder quarks
as compared to the three global sets.
As an illustration, the comparison of the gluon-gluon luminosity between the 
five NNLO PDF sets considered is shown in Fig.~\ref{fig:bench}.\footnote{An complete extensive
set of comparison plots can be found online at the NNPDF HepForge
website,\\
\url{https://nnpdf.hepforge.org/html/pdfbench/catalog/}
}

%%%%%%%%%%%%%%%%%%%%%%%%%%%%%%%%%%%%%%%%%%%%%%%%%%%%%%%%%%
%%%%%%%%%%%%%%%%%%%%%%%%%%%%%%%%%%%%%%%%%%%%%%%%%%%%%%%%%%

\paragraph{Constraints from LHC data.}

A major update since DIS2012 has been the inclusion of LHC data into
PDF analysis. 
Currently,  NNPDF2.3 is the only publicly available
PDF set that includes constrains from ATLAS, CMS and LHCb
jet and $W,Z$ data, but  other groups have presented
preliminary updates of their fits with the presence of LHC data.
For instance, ABM has studied the impact of LHC electroweak
boson data into their PDFs in~\cite{Alekhin:2013kla,Alekhin:2013dmy}.

The traditional processes at hadron colliders used for
PDF constraints are inclusive jets and $W,Z$ production.
Inclusive jet and dijet data are now available up to the TeV region
from ATLAS and CMS~\cite{Aad:2011fc,Chatrchyan:2012bja}, 
and provide important constrains on the poorly known large-$x$ quarks
and gluons.
On the other hand, the neutral current Drell-Yan processes has now been measured
both in the $Z$ peak region as well as for high and low
masses by ATLAS, CMS and LHCb, providing useful information
on the large-$x$ quarks and antiquarks (for high mass)
and in the small-$x$ gluon and possible departures from
lineal DGLAP evolution (for low mass DY).

\begin{figure}[h]
\centering
\includegraphics[scale=0.37]{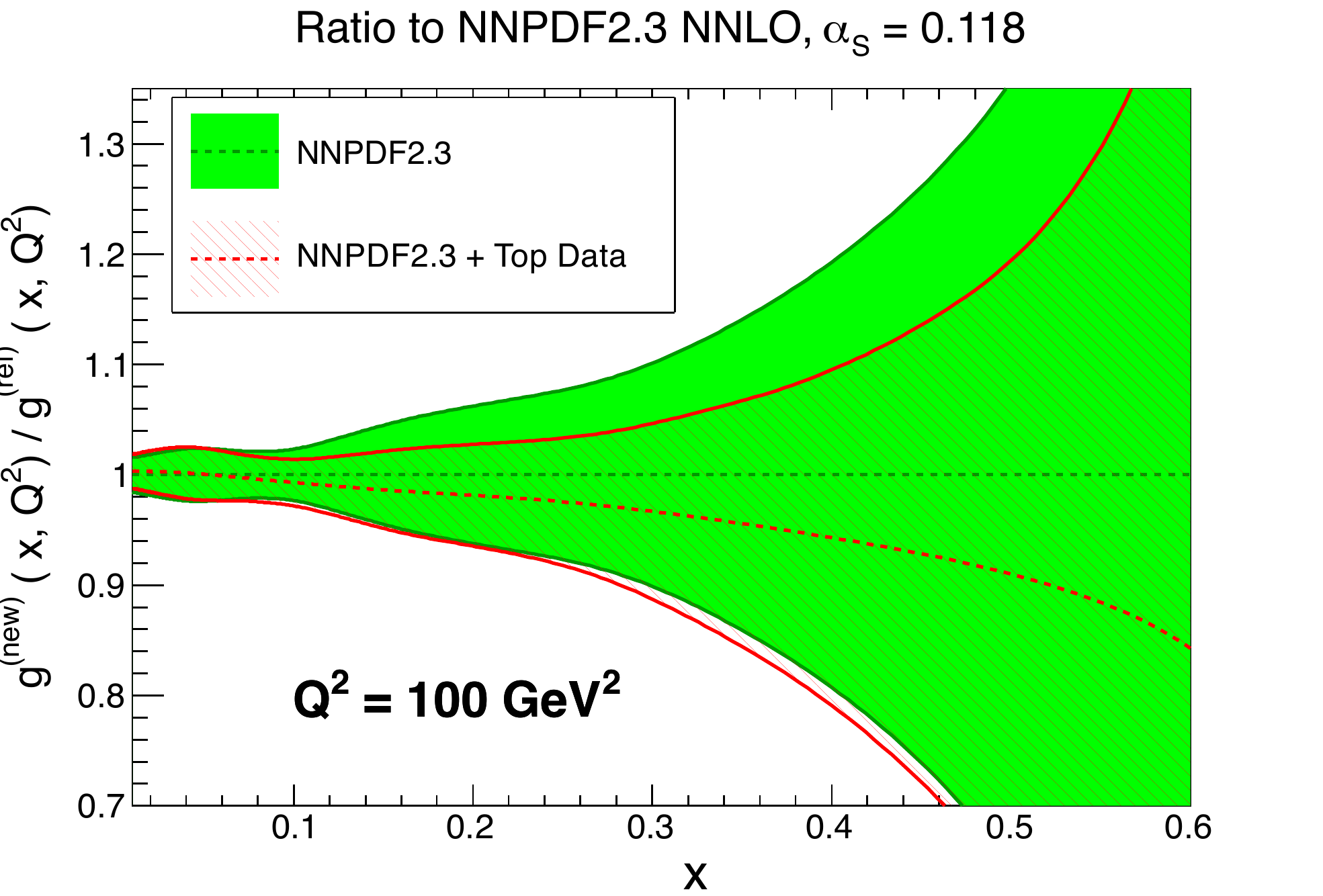}
\includegraphics[scale=0.18]{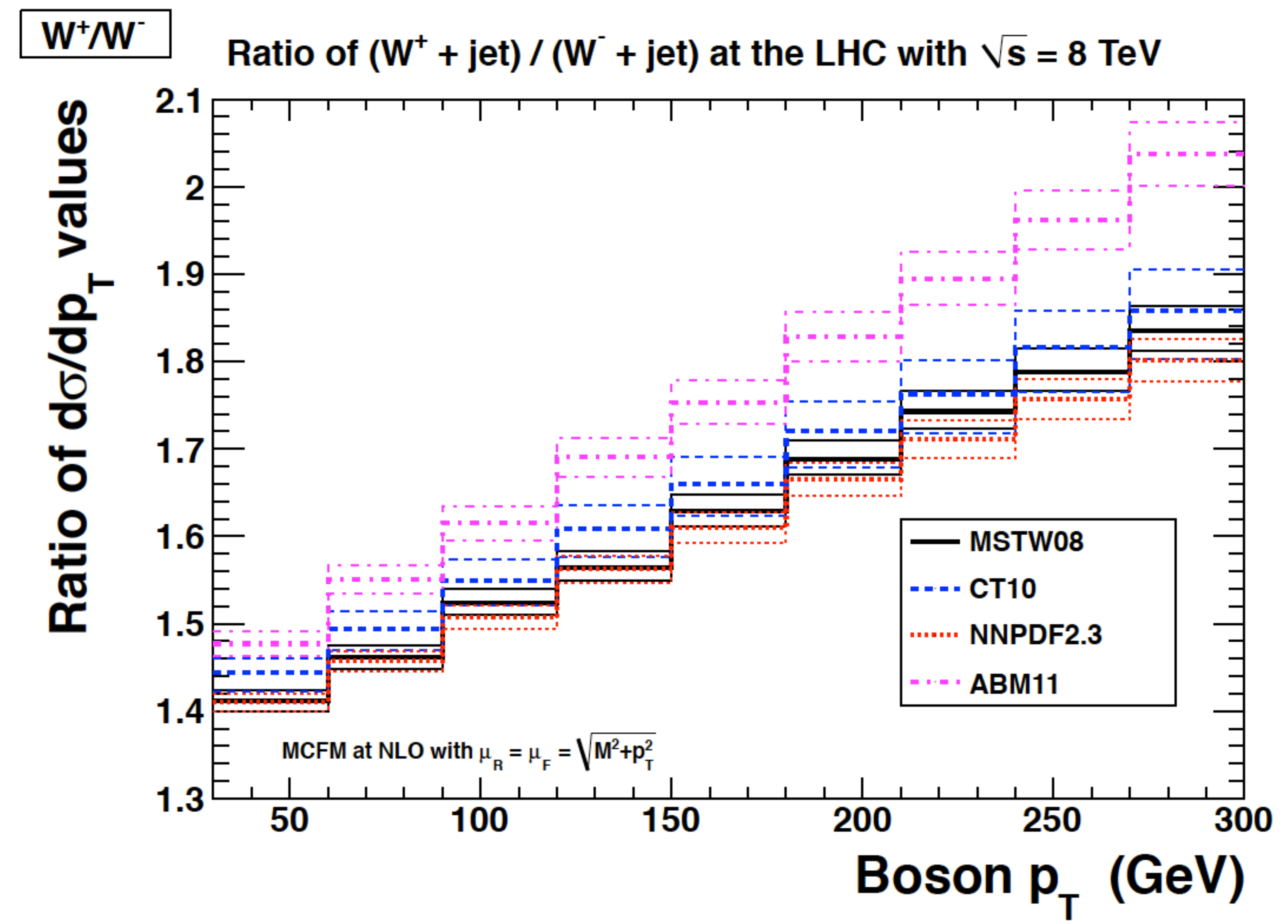}
\caption{\small  (left) The NNPDF2.3 NNLO gluon
at large-$x$ after the inclusion of LHC top
quark cross-section data~\cite{Czakon:2013tha}. (right) PDF sensitivity
of the ratio $W^+/W^-$ at high values of $p_T$~\cite{Malik:2013kba}.
}
\label{fig:data}
\end{figure}

In addition, thanks to the wealth
of LHC data several new processes have become available for their
use in PDF fits. 
The recent calculation of the full NNLO top quark production~\cite{Czakon:2013goa}
cross section makes  possible for the first time to include top quark
data into a NNLO analysis to constrain the large-$x$
gluon PDF~\cite{Czakon:2013tha} (see Fig.~\ref{fig:data}).
Top production is currently the only hadronic observable which is both directly
sensitive to the gluon and can be included in a  NNLO global fit
without any approximation.
In turn, the more accurate gluon PDF translates into
an improvement of the theory predictions for various high-mass
BSM processes driven by the gluon luminosity.
Another interesting result of Ref.~\cite{Czakon:2013tha} is that
the gluon determined from a  NNPDF2.1 fit to inclusive DIS HERA data
only~\cite{Ball:2011mu,Ball:2011uy} and
supplemented by LHC top production data is quite close
to that of the global fit, driven by jet data, a remarkable
consistency test of the global QCD analysis framework.
The use of LHC isolated photon data and photon+jet
data has also been advocated in order
to pin down the gluon PDF~\cite{d'Enterria:2012yj,Carminati:2012mm}, though 
this process is affect by missing higher order 
and non perturbative uncertainties.

%%%%%%%%%%%%%%%%%%%%%%%%%%%%%%%%%%%%%%%%%%%%%%%%%%%%%%%%%%
%%%%%%%%%%%%%%%%%%%%%%%%%%%%%%%%%%%%%%%%%%%%%%%%%%%%%%%%%%

Turning to the constraints on the quark sector,
the production of $W$ and $Z$ bosons
in association with jets, for high $p_T$ values
of the electroweak boson, is a clean probe at
the LHC of both quark flavor 
separation and of the gluon PDF~\cite{Malik:2013kba}.
In particular, ratios of $W$ and $Z$ distributions at large $p_T$
provide constrains on quarks and antiquarks while benefiting from
substantial cancellations of experimental and theory uncertainties, see
Fig.~\ref{fig:data}.
Another important source of information is 
$W$ production with charm quarks, directly sensitive to the strange
PDF~\cite{Stirling:2012vh}, the worst known of all light
quark flavors. This process  has been recently measured with CMS~\cite{CMSWc}, showing
good agreement with the strangeness suppression of global
PDF fits derived from the neutrino DIS charm production data,
while ATLAS~\cite{ATLASWc} prefers a symmetric strange sea, similar as 
previously derived from their inclusive $W$ and $Z$ data.
Including all these datasets into the 
  global PDF fits would then be crucial to determine the optimal
  strange PDF which accounts for all experimental constraints.

A recently proposed useful observable at the LHC that can be used
to enhance the sensitivity to PDFs are cross section
ratios between different beam energies. As discussed 
in~\cite{Mangano:2012mh}, the advantage of these ratios is that
many theoretical systematics cancel, while at the same
time the sensitivity to PDFs is maintained. 
In addition, many experimental
uncertainties cancel in a dedicated measurement of a cross section
ratio, so the net result is a very clean probe of differences
between PDF sets.
A first implementation of this idea has been presented by ATLAS,
with the measurement the ratio of inclusive jet cross sections
between 2.76 TeV and 7 TeV~\cite{Aad:2013lpa}.
The measurement of cross section ratios between 14 and 8 TeV will provide further
direct constrains on various PDF combinations, as well as being
potentially useful to enhance the sensitivity to BSM contributions as
compared to absolute cross sections~\cite{Mangano:2012mh}.

The list of additional LHC processes that can potentially used for PDF studies includes
single top, open heavy quark and charmonium production among others. 
Is clear that within a few years PDFs based only on collider data only might
become competitive with global PDF sets.

\paragraph{Theoretical developements.}

The theory frontier in PDF determinations is NNLO, however
up to recently only a  small number of hadron collider
processes relevant for PDFs were available at this order,
essentially electroweak vector 
boson production~\cite{Anastasiou:2003ds,Catani:2010en,Gavin:2010az}. 
In the last year, substantial progress has been made in this direction:
the full NNLO top quark production cross section was 
computed~\cite{Czakon:2013goa}, the NNLO corrections to jet production
in the $gg$ channel became available~\cite{Ridder:2013mf}, and the Higgs plus 
one jet was completed~\cite{Boughezal:2013uia}, which suggests that
the NNLO calculation for W or Z in association jet might become available
in the near future. 

\begin{figure}[h]
\centering
\includegraphics[scale=0.36]{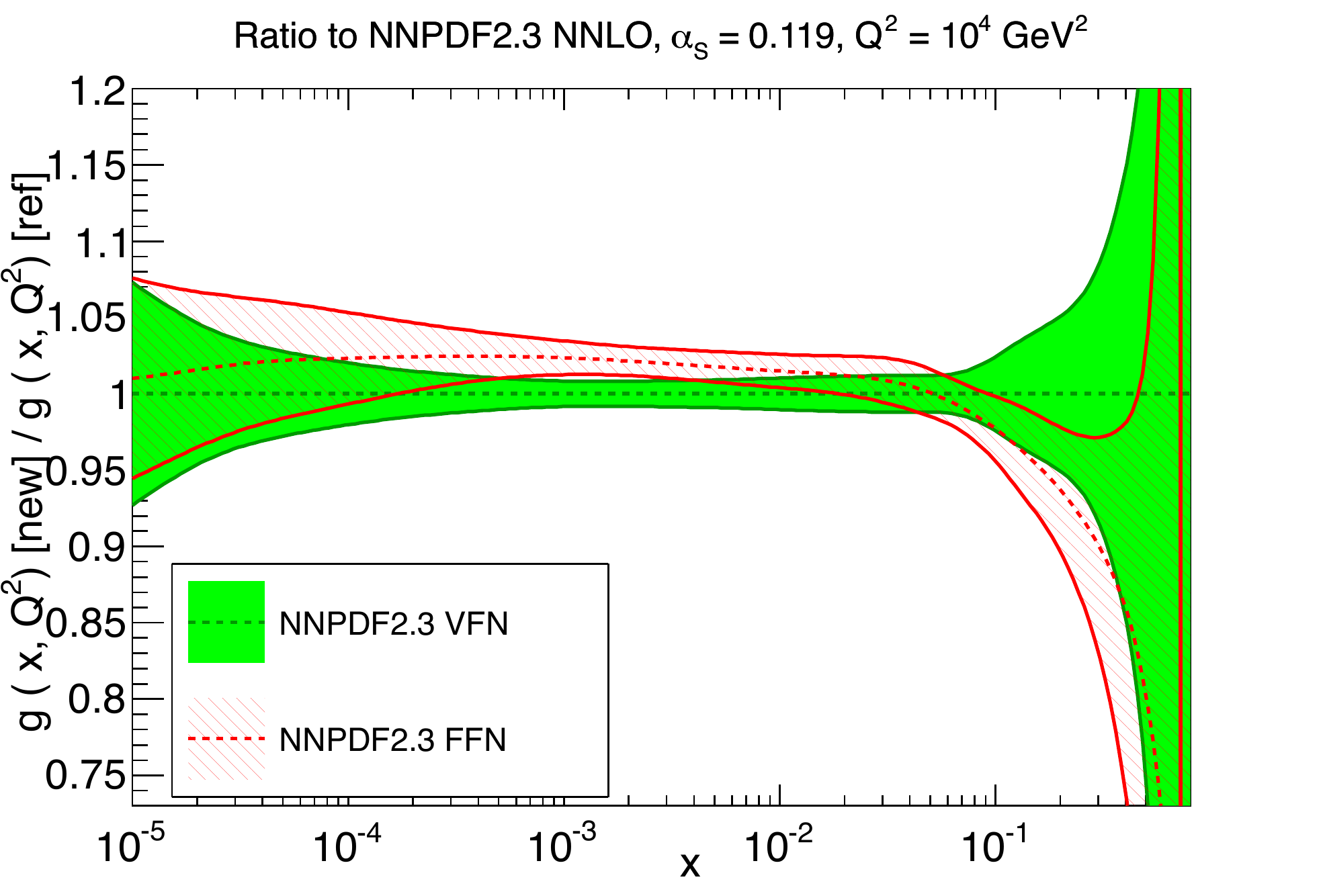}
\includegraphics[scale=0.14]{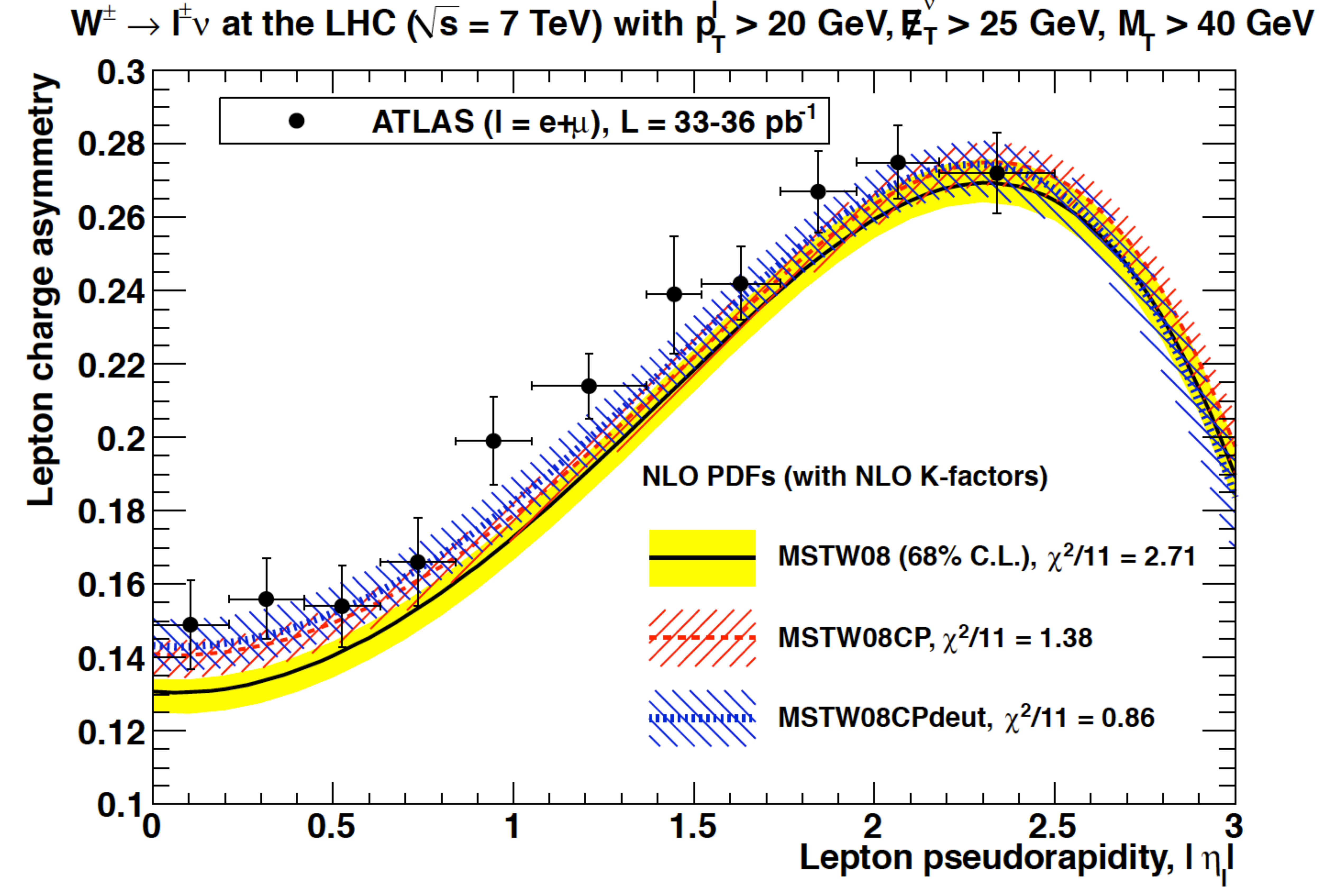}
\caption{\small  (left) The ratio  of gluon 
PDF at $Q^{2}=10^4$ GeV$^2$ extracted using 
VFN and FFN renormalization schemes in the NNPDF2.3 analysis. (right) The improved description
of the LHC $W$ asymmetry data in the MSTW analysis due to
the more flexible Chebyshev input PDF parametrization, from~\cite{Martin:2012xx}. 
}
\label{fig:meth}
\end{figure}

While the consequences of experimental uncertainties in PDF fits
have been studied extensively, less work has been devoted
to the implications of theoretical uncertainties.
Recently, a number of studies have become available 
that attempt to quantify the role of various
potential sources of theoretical uncertainties in the determination
of the parton distributions. 
To begin with, 
the impact of different choices of  quark flavor number renormalization
schemes in fitting DIS data has been explored by NNPDF~\cite{Ball:2013gsa} 
and by Thorne~\cite{Thorne:2012az}.
These studies consistently indicate that the fixed-flavor number scheme leads to
softer large-$x$ gluons and harder quarks as compared to
PDFs determined in general-mass variable flavor number schemes (GM-VFN),
 suggesting that the use of different  quark flavor number renormalization
 schemes (GM-VFN versus FFN) might  account for
part of the differences between ABM11 and global PDF fits.
As an illustration, we show in Fig.~\ref{fig:meth} the
NNPDF2.3 gluon at a typical LHC scale determined in a fit
with FFN scheme, as compared to the default fit in the FONLL-C GM-VFN scheme~\cite{Forte:2010ta}.
In addition, these studies tend to indicate that a somewhat poorer fit quality
to the large $Q^2$ HERA data is obtained in the FFN as compared to VFN,
perhaps related to the missing resummations of DGLAP
logarithms~\cite{Caola:2010cy}.

The sensitivity of the global PDF fit to the $W^2$ cut, higher twists 
and deuteron corrections
 has been also studied by NNPDF and MSTW~\cite{Martin:2012xx}.
The main conclusions are that while with standard $W^2$ cuts the impact of higher twists is negligible,
deuteron corrections lead to moderate modifications of the $d/u$ ratio, but restricted
to the range $0.1\lsim x \lsim 0.5$.
The  CJ12 
analysis~\cite{Owens:2012bv} has also studied the treatment
of deuteron corrections and the role of higher-twist effects which are
important at small $W^2$.
%

%%%%%%%%%%%%%%%%%%%%%%%%%%%%%%%%%%%%%%%%%%%%%%%%%%%%%%%%%%
%%%%%%%%%%%%%%%%%%%%%%%%%%%%%%%%%%%%%%%%%%%%%%%%%%%%%%%%%%

Another area of theoretical active developement 
are PDFs with QED corrections. Precision
electroweak computations at hadron colliders require PDFs
with a combined QCD and QED evolution, and in turn this requires  the
determination of the photon PDF in the proton.
Photon-initiated corrections are relevant for the determination of $M_W$,
for background estimates of high mass $Z'$ and $W'$ searches and
for other electroweak processes such as $WW$ production at large invariant mass.
 The only available set
up to know is MRST2004QED~\cite{Martin:2004dh}, where the 
initial photon parametrization is derived
from a QED radiation inspired model.
 Both NNPDF and CT have
presented preliminary results towards updated PDF sets with
combined QCD and QED corrections. 
In the NNPDF case~\cite{sc}, the photon PDF is based on the standard artificial
neural networks parametrization, without any non-perturbative
assumption, and fitted to DIS and LHC electroweak gauge boson production
data. 
In constraining the photon PDF, the direct photon production data from
the HERA collider can also be useful~\cite{Chekanov:2009dq}.

As a last theory issue,
the accurate treatment of heavy quark mass effects is important
in the fits to HERA cross-section data.
The role of heavy quark masses in the description of HERA charm
production data has
been recently discussed by ABM~\cite{Alekhin:2012vu}, 
CT~\cite{Gao:2013wwa} and HERAPDF~\cite{Abramowicz:1900rp}, 
where the running
charm mass $m_c\lp m_c \rp$ has been extracted from the combined
HERA dataset with different treatment of heavy quark
structure functions, and the implications for $W$ and $Z$ production
at the LHC have been discussed. 
These results are in general
in good agreement with the PDG value.

\paragraph{Methodological studies.}

\begin{figure}[h]
\centering
\includegraphics[scale=0.43]{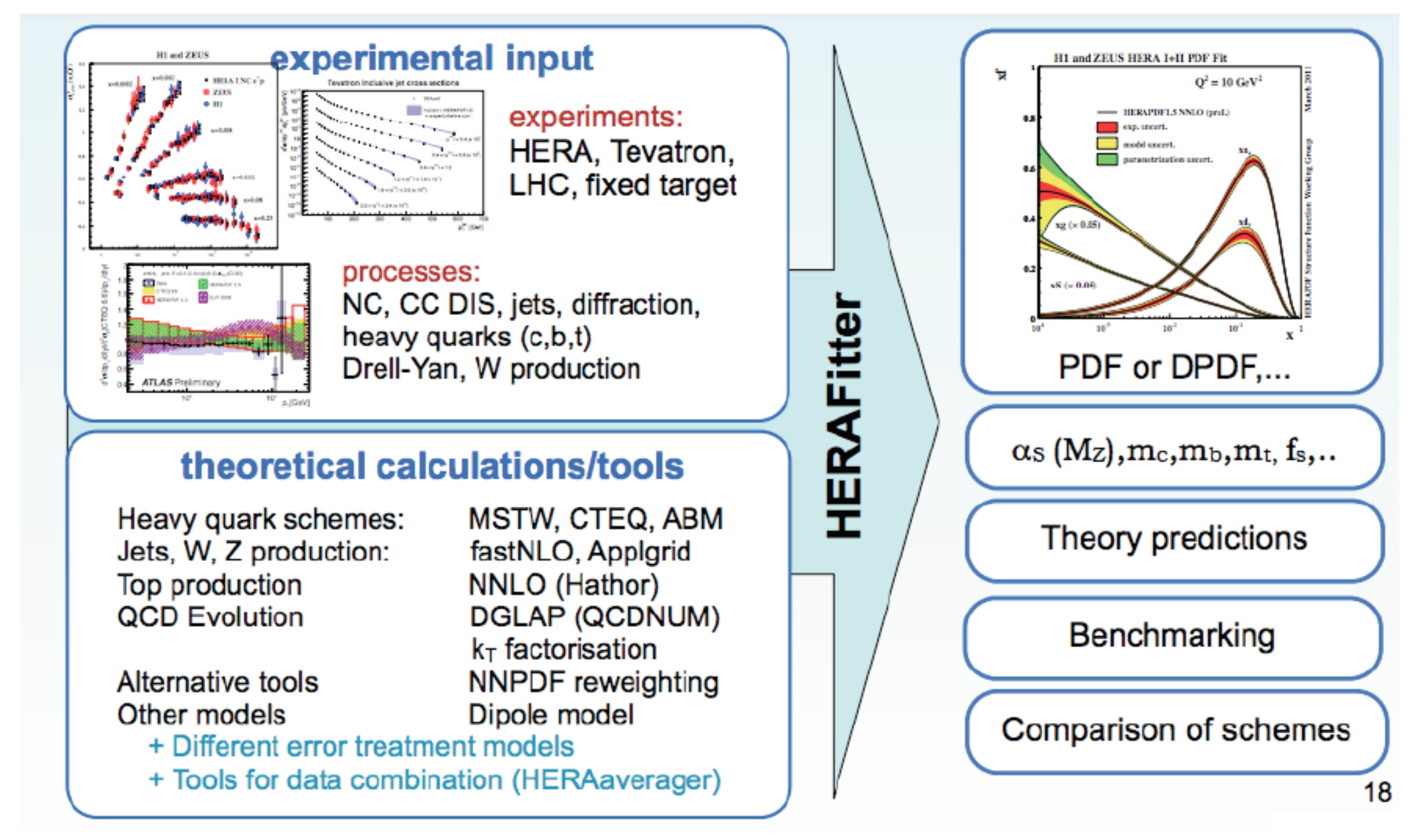}
\caption{\small  (left) Summary of the {\tt HERAfitter} project.}
\label{fig:herafitter}
\end{figure}

Moving to methodology,
an important recent development is that of the {\tt HERAfitter} framework,\footnote{\url{https://www.herafitter.org/HERAFitter/}}
an open-source tool for PDF analysis, whose aim is to provide
a flexible code, open to contributions from all interested parties,
that is able to perform PDF fits, quantify the impact of
new data, perform $\chi^2$ tests and study the dependence on
various theory choices among others.
{\tt HERAfitter}  has been used in various experimental analysis
by HERA, ATLAS and CMS, and will provide important input
both from the theoretical and experimental point of view
to global PDF fits.
A summary of the {\tt HERAfitter}  project is shown in Fig.~\ref{fig:herafitter}.

%%%%%%%%%%%%%%%%%%%%%%%%%%%%%%%%%%%%%%%%%%%%%%%%%%%%%%%%%%
%%%%%%%%%%%%%%%%%%%%%%%%%%%%%%%%%%%%%%%%%%%%%%%%%%%%%%%%%%

As discussed in the PDF benchmark comparison of Ref.~\cite{Ball:2012wy}, MSTW08 provides
a poor description of the $W$ asymmetry data from ATLAS and CMS.
This problem can be traced back to a not general enough parametrization
of the $u_V-d_V$ PDF combination.
In an update of their global fit, MSTW has presented a more flexible parametrization of their input PDFs
based on Chebyshev polynomials~\cite{Martin:2012xx}.
 These extended
parametrizations lead  to an improved description of the LHC $W$ asymmetry data
starting from the same  non-LHC dataset.
This improved description, 
 shown in Fig.~\ref{fig:meth}, is related  to a modification
in the $u_V-d_V$ combination that arises when the Chebyshev polynomials are used.

Another important achievement from the methodology point of view is 
the generalization of the PDF reweighting
methods, originally developed for Monte Carlo
PDF sets~\cite{Ball:2011gg,Ball:2010gb}, to the case
of Hessian based PDFs~\cite{Watt:2012tq}. 
These reweighting methods allow to efficiently quantify the impact
of new data into PDFs without the need of refitting, for instance, can
be used by the experimental collaborations.
Therefore, Bayesian PDF reweighting can now be applied to
all available PDF sets, both Hessian and Monte Carlo.

One important aspect of the statistical framework of PDF
analysis is that
various definitions of the figure of
merit $\chi^2$ exist in the literature, leading
to different results when used in the PDF fit.
Statistical issues related to these definitions of the $\chi^2$ have been discussed recently, see
for instance Appendix A of ~\cite{Ball:2012wy}.
 While different possible
treatment of multiplicative uncertainties are possible, only
the so-called $T_0$ method~\cite{Ball:2009qv} is strictly
unbiased. 
Note that not only normalization errors are multiplicative, many other experimental
systematics, like the jet energy scale, should also be treated as such.
In this respect, the CT collaboration
 has studied the dependence of the gluon PDFs
on various possible $\chi^2$ definitions of the inclusive jet data~\cite{Gao:2013xoa}, showing
that at large-$x$ the different $\chi^2$ can lead to modifications comparable
to the impact of jet data itself.
Related issues have also been studied by ABM~\cite{Alekhin:2012ce}
and earlier by Thorne and Watt~\cite{Thorne:2011kq}.

As a last remark concerning methodological
improvements, let us mention that in the first NNPDF polarized analysis, {\tt NNPDFpol1.0}, the determination
of the effective large and small-$x$ PDF exponents has been
studied~\cite{Ball:2013lla}.
These effective exponents
 provide  useful information
to compare the limiting PDF behaviour with non-perturbative models, without the need to
introduce an ad-hoc polynomial parametrization.

\paragraph{Theoretical speculations.}
In the last part of  this contribution, let me speculate some possible
future directions for possible
theory developements in PDF analysis in the coming years:
\begin{itemize}
\item Do we need PDFs at (approximate) N3LO, to match the accuracy
of foreseen Higgs cross section calculations? 
\item Do we need PDFs supplemented with threshold resummation,
to match the accuracy of processes like Higgs and top production
were NNLO+NNLL is the accuracy frontier?
\item We need to include electroweak corrections in the fit
to the TeV LHC data? Is it enough to include them at the matrix
element level or some modification of the DGLAP evolution
might be needed?
\item Do we need specific PDFs for NLO event generators at the
LHC? What is the difference between NLO and NLO+PS PDFs?
  Is it possible to provide a PDF set that provides a simultaneous
description of hard scattering and semihard data?
\item Are potential intrinsic heavy flavours relevant for
LHC phenomenology?
\item Can we design new avenues in using the global QCD framework
to derive useful constrains on BSM dynamics from LHC data?
\end{itemize}

\paragraph{Outlook.}
Driven by the needs of precision LHC physics, there has been
substantial progress in PDF determinations in the last year.
PDFs are already an important ingredient of the 
LHC physics program,  essential for the characterization
of the Higgs boson, and required for the understanding of  potential New Physics signals that the LHC hopes to
find after the energy upgrade in 2015.

\providecommand{\href}[2]{#2}\begingroup\raggedright\endgroup

\end{document}